\documentclass[aip,twocolumn,superscriptaddress]{revtex4}

\usepackage{graphicx}
\usepackage{bm} 

\begin{document}


\title{Micro-magnets for coherent control of spin-charge qubit in lateral quantum dots}


\author{M. Pioro-Ladri\`{e}re}
\affiliation{Quantum Spin Information Project, ICORP, Japan Science and Technology Agency, Atsugi-shi, Kanagawa, 243-0198, Japan}

\author{Y. Tokura}
\affiliation{Quantum Spin Information Project, ICORP, Japan Science and Technology Agency, Atsugi-shi, Kanagawa, 243-0198, Japan}
\affiliation{NTT Basic Research Laboratories, NTT Corporation, Atsugi-shi, Kanagawa, 243-0198, Japan}

\author{T. Obata}
\affiliation{Quantum Spin Information Project, ICORP, Japan Science and Technology Agency, Atsugi-shi, Kanagawa, 243-0198, Japan}

\author{T. Kubo}
\affiliation{Quantum Spin Information Project, ICORP, Japan Science and Technology Agency, Atsugi-shi, Kanagawa, 243-0198, Japan}

\author{S. Tarucha}
\affiliation{Quantum Spin Information Project, ICORP, Japan Science and Technology Agency, Atsugi-shi, Kanagawa, 243-0198, Japan}
\affiliation{Department of Applied Physics, University of Tokyo, Hongo, Bunkyo-ku, Tokyo, 113-0033, Japan}


\date{\today}

\begin{abstract}
A lateral quantum dot design for coherent electrical manipulation of a two-level spin-charge system is presented. Two micron-size permanent magnets integrated to high-frequency electrodes produce a static slanting magnetic field suitable for voltage controlled single qubit gate operations. Stray field deviation from the slanting form is taken into account in the Hamiltonian describing the two-level system, which involves hybridization of a single electron spin to the quantum dot's orbitals. Operation speed and gate fidelity are related to device parameters. Sub 100 ns $\pi$ pulse duration can be achieved with lattice fluctuations coherence time of 4 ms for GaAs.
\end{abstract}


\maketitle


Quantum dots (QDs) are man-made structures which can confine conduction electron in semiconductor to a nanometer size volume \cite{Kouwenhoven98,Jacak98}. In the few-electron regime, the spin of individual electrons, which is a natural two-level system, can be used as a quantum bit for implementing scalable quantum computing \cite{Loss98}. In this context various experiments have been performed on robustness of electron spin in single- and double-dot systems \cite{Fujisawa01,Petta05}, spin correlation \cite{Ciorga02,Sasaki04,Hatano05}, electrical readout of single electron spin \cite{Elzerman04} and coherent manipulation of two-electron spin states  \cite{Petta05sci}. The rotation of a single electron spin or qubit operation, which is the most fundamental quantum operation, was recently demonstrated by generating an on-chip electron spin resonance (ESR) magnetic field \cite{Koppens06}. An alternative approach is to modulate a QD electric field in a non-uniform magnetic field \cite{Tokura06}. This scheme eliminates the need for an externally applied ac magnetic field, and does not require spin-orbit coupling, as opposed to earlier work on electron spin control based on g-tensor modulation, and on electric fields \cite{Kato03,Rashba03}. Instead, ESR is achieved by applying microwave gate voltage pulses, letting the electron position oscillate in a static slanting Zeeman field. This effectively provides the electron spin with the necessary time-dependent transverse magnetic field. 
  
We present in this paper a lateral quantum dot design to perform electrical ESR in a slanting Zeeman field. The design involves a realistic permanent magnets configuration to produce the required inhomogeneous static magnetic field and high-frequency electrodes for fast charge control. The geometry used is that of well-established few-electron quantum dots defined in semiconductor two-dimensional electron gas (2DEG) by surface gates \cite{Ciorga00}. The coupling between the spin and orbital degrees of freedom allows fast qubit rotations with decoherence time only slightly reduced from the intrinsic spin dephasing time for GaAs quantum dots.

\begin{figure}
\includegraphics[scale=1.0]{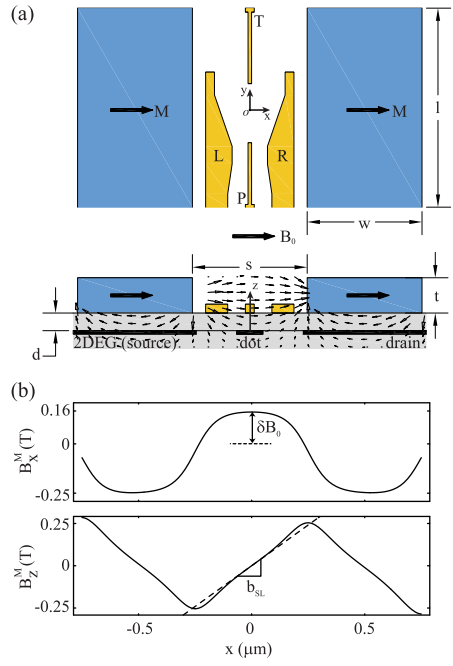}
\caption{\label{device} (a) Top and cross-sectional views of the lateral quantum dot device consisting of four  metallic gates (yellow) and two ferromagnetic strips (blue) patterned at the surface of a semiconductor heterostructure. The thick arrows indicate the direction of the external magnetic field $B_{0}$ and magnetization $M$. The origin is fixed to the quantum dot position. (b) In-plane ($B^{M}_{x}$) and transverse ($B^{M}_{z}$) profile of the stray magnetic field produced for two Co micromagnets ($\mu_{0}M=1.8\,\mathrm{T}$) calculated at $z=0$. Dimensions $d$, $t$, $s$, $w$ and $l$ are set to $0.1$, $0.2$, $0.5$, $0.5$ and $2.0\,\mathrm{\mu m}$.}
\end{figure}

The device layout is presented in Fig. \ref{device}.a. Four metallic gates, located on the surface of a semiconductor heterostructure, form Schotky contacts to a 2DEG located at a distance $d$ below. For GaAs/AlGaAs wafer, $d$ is typically 100 nm. By applying suitable dc gate bias, a quantum dot confining a single excess electron, separated from source and drain reservoirs by tunnel barriers, is formed in the 2DEG. The QD confining potential is composed of a transverse part $V\left(z\right)$ coming from the band structure and in-plane part $V\left(x,y\right)$ electrostatically defined by gates L, P, R and T. For typical gate bias, the confinement can be modeled as $V(x,y)=\frac{1}{2}m\omega_0^2\left[x^2+y^2+\gamma(y-\frac{x^2}{D})^2\right]$ \cite{Kyriakidis02} where $\gamma$ characterizes nonparabolicity and the length $D$ is the diameter of semicircular confinement by the T gate. The slanting magnetic field is produced by two ferromagnetic strips positioned on each side of the gate structure (see below). Each strip is connected to a high-frequency voltage port through impedance-matched on-chip waveguide to enable $\mathrm{GHz}$ electric field modulation. Anti-phase modulation of the strips potential at frequency $\nu$ produces a uniform electric field $\vec{E}=E_{0}\sin\left(2\pi\nu t\right)\hat{x}$ inside the quantum dot with amplitude $E_0$ proportional to the potential drop $V_0$ between the strips and inversely proportional to the strip separation $s$. From numerical simulations of the microwave field produced by the strips, we find $E_0 = 7.71\times10^{-7}\,V_{0}/s$ in S.I units. 

The external magnetic field, required for the ESR scheme, is applied in the plane of the QD whose strength $B_{0}$ (few Teslas) is much higher than the coercive field of the patterned magnets (ranging from zero to hundreds of Gauss). In these conditions, the strips are magnetized uniformly in the direction parallel to the external field. The magnetic moment per unit volume is equal to the saturation magnetization $M$ of the ferromagnetic material. The resulting stray magnetic field is illustrated in Fig. \ref{device}.a. In the magnets plane ($z=d+t/2$), the stray field is parallel to the external field. In the 2DEG plane ($z=0$) however, a perpendicular component is present due to the curvature of the field lines. Figure \ref{device}.b shows the profiles of the parallel ($B^{M}_{x}$) and transverse ($B^{M}_{z}$) components of the stray field at the 2DEG location. The transverse component peaks with opposite sign at the strips edges $x=\pm g/2$ while the in-plane component switches sign at these positions. These properties produce, near the QD location, a total magnetic field of the form
\begin{widetext}
\begin{equation}	\vec{B}=\left[B_0+\delta B_{0}+b_{SL}z+a\left(z^{2}-x^{2}\right)+3bzx^{2}-bz^{3}\right]\hat{x} + \left[b_{SL}x+2axz-3bxz^{2}+bx^{3}\right]\hat{z}
	\label{B}
\end{equation}
\end{widetext}
which satisfies Maxwell's equations $\vec{\nabla}\cdot\vec{B}=0$ and $\vec{\nabla}\times\vec{B}=0$. 

The terms in Eq. (\ref{B}) up to the first order in $x$ and $z$ displacements correspond to the slanting form previously considered in Ref. \cite{Tokura06}. These terms are described by a uniform shift $\delta B_{0}$ of the external field and gradients with slope $b_{SL}$ on the order of $1\,\mathrm{T/\mu m}$. Coefficients $a$ and $b$ give respectively the second and third order corrections to the slanting form. For displacement smaller than $\pm 25\,\mathrm{nm}$, we find that Eq. (\ref{B}) fits very well the magnetic field profile obtained by numerical simulation \footnote{We used Mathematica Radia package available at http://www.esrf.fr/ for stray field calculations.}, with departure of less than $0.1\%$. Because of the two-dimensional nature of the QD (i.e. strong transverse confinement), the effect of the $z$ terms in Eq. (\ref{B}) on the electron dynamics can be averaged out. This procedure involves averages $\left\langle z^k \right\rangle=\left\langle \xi_{0}\right|z^k\left|\xi_{0}\right\rangle$ over the ground state wavefunction $\xi_{0}\left(z\right)$ of $V\left(z\right)$. Term $k=1$ is equal to zero and terms $k=2,3$ renormalize the shift and slope to $\delta \tilde{B}_0=\delta B_{0} + a \left\langle z^{2}\right\rangle - b\left\langle z^{3}\right\rangle$ and $\tilde{b}_{SL}= b_{SL}-3b\left\langle z^{2}\right\rangle$ to give the effective QD magnetic field $\left\langle\vec{B}\right\rangle=\left[{B}_{0}+\delta\tilde{B}_{0}-ax^{2}\right]\hat{x}+\left[\tilde{b}_{SL}x+bx^{3}\right]\hat{z}$. 

We now relate the proposed device to the control of the hybrid spin-charge qubit. With $M=0$, the electron spin confined into the QD is described by the Hamiltonian $H_0=E_0 +\left(p_{x}^{2}+p_{y}^{2}\right)/2m + V\left(x,y\right)-g\mu_{B}B_0\sigma_x/2$ where $E_0$ is the ground state energy of the transverse Hamiltonian, $m$ the effective mass, $g$ the g-factor and $\mu_B$ the Bohr magneton. Throughout the text, we use material parameters of GaAs: $m=0.067m_{e}$ and $g=-0.44$ where $m_{e}$ is the electron mass. The Pauli spin matrices $\vec{\sigma}=\left(\sigma_{x},\sigma_{y},\sigma_{z}\right)$ are decoupled from the QD orbitals $\phi_{n}\left(x,y\right)$. The eigenenergies and eigenfunctions of $H_0$ are $\epsilon_{n\sigma}=E_{0}+\epsilon_{n}-\frac{1}{2}g\mu_{B}B_{0}\sigma$ with eigenfunctions $\left\langle x,y,z | n,\sigma \right\rangle=\xi_{0}\left(z\right)\phi_{n}\left(x,y\right)\psi_{\sigma}$ where $n=0,1,2,...$, $\sigma=\pm 1$ and $\psi_{\sigma}$ is the spinor of $\sigma_{x}$ eigenstates. The ground state wave function is spin split by the Zeeman energy which is assumed to be smaller than orbital excitation energy: $g\mu_{B}B_{0}<\epsilon_{1}-\epsilon_{0}$.

With $M>0$, the Hamiltonian becomes $H=H_0+W_{x}\left(x\right)\sigma_x+W_{z}\left(x\right)\sigma_{z}$ with position dependent perturbations $W_{x}\left(x\right)=-g\mu_{B}\left(\delta \tilde{B}_{0}-ax^2\right)/2$ and $W_{z}\left(x\right)=-g\mu_{B}\left(\tilde{b}_{SL}x+bx^{3}\right)/2$. The relatively large value of $\delta \tilde{B}_{0}$ shifts the Zeeman energy of the QD spin from the nearby reservoirs, a useful feature for read-out schemes based on spin to charge conversion \cite{Engel01}. Terms $W_{x,z}\left(x\right)$ mix the electron spin states $\left|\sigma\right\rangle$ with the orbital states $\left|n\right\rangle$. For example, the two lowest energy states, which constitute our "hybridized" spin-charge qubit, are, for the harmonic potential ($\gamma = 0$), $\left|g\sigma\right\rangle\approx C_{0\sigma}\left|n_x=0,\sigma\right\rangle + C_{1-\sigma}\left|n_x=1,-\sigma\right\rangle$ where $n_x=0,1$ corresponds to the first two orbitals of the one dimensional harmonic potential (of the $x$ coordinate) \cite{Tokura06}. For the magnetic field profile shown in Fig. 1.b and typical confinement energy $\hbar\omega_{0}=1\,\mathrm{meV}$, we find $C_{1-\sigma}/C_{0\sigma}\approx0.0003$. The mixing between the spin and orbital states is therefore relatively small. Neglecting for the moment possible misalignment between the magnetic strips and the quantum dot, only the off-diagonal matrix elements of the high-frequency QD electric potential operator $-eE_{0}x\sin\left(2\pi\nu t\right)$ remain in the subspace spanned by $\left\{\left|g\sigma=+1\right\rangle,\left|g\sigma=-1\right\rangle\right\}$. The Hamiltonian of the qubit is thus of the ESR form
\begin{eqnarray}
\label{HESR}
H_{ESR}&=&\frac{1}{2}h\nu_{01}\hat{\sigma}_z+\frac{1}{2}\epsilon_{x}\sin(2\pi\nu t)\hat{\sigma}_x
\end{eqnarray}
where $h$ is the Planck constant and $\hat{\sigma}_x$ and $\hat{\sigma}_x$ are the Pauli matrices of the effective two-level system. 

\begin{figure}
\includegraphics[scale=1.0]{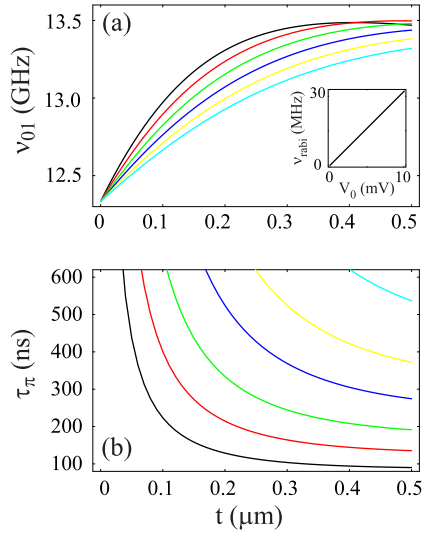}
\caption{\label{tdep} Calculated dependence of (a) spin-charge qubit's Larmor frequency $\nu_{01}$ and (b) gate operation speed $\tau_{\pi}$ for ESR voltage $V_0=1\,\mathrm{mV}$ on magnetic strips's thickness $t$ and separation $s$. Black, red, green, blue, yellow and magenta curves correspond to value $s$ of $0.5$, $0.6$, $0.7$, $0.8$, $0.9$ and $1.0\,\mathrm{\mu m}$ respectively. Other dimensions are same as in Fig. \ref{device}. Magnetization $\mu_{0}M=1.8\,\mathrm{T}$ is assumed. Inset: Rabi frequency $\nu_{\mathrm{rabi}}=1/\left(2\tau_{\pi}\right)$ as a function of $V_0$ for $t=0.2\,\mathrm{\mu m}$ and $s=0.5\,\mathrm{\mu m}$.}  
\end{figure}

The Larmor frequency $\nu_{01}$ ($h\nu_{01}$ corresponding to the qubit energy level separation) is slightly smaller than unperturbed Zeeman energy $g\mu_{B}(B_0+\delta\tilde{B}_{0})$. The ESR transverse field $\epsilon_{x}$, which controls qubit rotation speed, is proportional to the modulation voltage $V_0$ and magnetization $M$. By adjusting the phase of the modulation, arbitrary qubit rotations can be achieved when $\nu=\nu_{01}$. The predicted dependence of $\nu_{01}$ and $\pi$ pulse duration time, $\tau_{\pi}=h/\epsilon_{x}$, on strips's thickness $t$, separation $s$ and $V_0$ is presented in Fig. \ref{tdep}. Results are obtained by least square fit of the simulated stray magnetic field and microwave electric field followed by exact diagonalization of the corresponding spin-charge Hamiltonian $H$ for $B_0=2\,\mathrm{T}$. In the calculation, we use confining potential parameters $\hbar \omega_0=1\,\mathrm{meV}$, $\gamma=1$, and $D=2.9\sqrt{\hbar/m\omega_0}$ and magnetization of cobalt $\mu_{0}M=1.8\,\mathrm{T}$. For a given strip separation, the thicker are the strips, the bigger are $\nu_{01}$ and $\epsilon_{x}$. For thickness greater than approximately $4d$, $\nu_{01}$ and $\tau_{\pi}$ start to saturate to their $t\rightarrow\infty$ value. Only small improvement is therefore achieved in using strips thicker than $0.4\,\mathrm{\mu m}$. Small $s$ should be used as it improves drastically qubit rotation speed. For relatively small electric field $E_0\approx5\,\mathrm{mV/\mu m}$ (corresponding to point $V_0=3\,\mathrm{mV}$ in inset of Fig. 2a), the rotation speed is comparable to the fastest Rabi oscillations achieved recently in GaAs lateral quantum dot with on-chip ESR magnetic coil operating at much lower frequency ($\nu=200\mathrm{MHz}$) \cite{Koppens06}. We speculate that for the GHz regime, Rabi oscillations faster than with conventional ESR should therefore be observed. The misalignment between the strips and the QD introduces an oscillating term $\epsilon_{z}\sin\left(2\pi\nu t\right)\hat{\sigma}_z$ in the qubit Hamiltonian. For misalignment as large as $100\,\mathrm{nm}$, we find that this term does not influence gate operations, i.e. the gate fidelity for $\pi$ pulse remains one with error less than $10^{-7}$.

Usual spin qubit, in the absence of spin-orbit coupling, are naturally protected against lattice fluctuations. For our case, because the spin states are coupled to the electron's orbitals, effect of phonons has to be considered. As done by Tokura {\it et al}. \cite{Tokura06} for one-dimensional systems, acoustic phonon scattering causes orbital relaxation between the qubit states. Slightly modifying the previous analysis to our two-dimensional quantum dot, we estimate a relaxation time $T_1\sim 2.1\,\mathrm{msec}$ at $B_0=2\,\mathrm{T}$, which is dominated by transverse piezoelectric scattering. Since the dephasing effect without relaxation is negligible, the coherence time is $2T_1$. The corresponding quality factor $Q$ is estimated by coherence time divided by $\tau_{\pi}$, which is order of $10^4$, similar to the one-dimensional system considered previously. $Q$ factor of the same order has been estimated for the case of spin-orbit coupling \cite{Nazarov01}. The main source of decoherence comes from the hyperfine interaction with GaAs nuclei spins. The nuclei will induce fast dephasing (with time averaged coherence time of 10-20 ns) as demonstratred by the work of Petta {\it et al} \cite{Petta05sci}. However, because the hyperfine field fluctuates on a slow time scale compared to gate operation speed, this effect can be corrected using compensation \cite{Petta05sci,Taylor2005} or projection techniques \cite{Loss2006}.

We acknowledge W.G. van der Wiel for useful discussions and T. Taniyama and Y. Sekine for inputs on micro-magnets fabrication. ST thanks financial supports from the DARPA (DAAD19-01-1-0659) of the QuIST program, the Grant-in-Aid for Scientific Research A (No.
40302799), SORST-JST and IT Program, MEXT.

\newpage


\newpage

\end{document}